\documentstyle[12pt,epsf]{article}
\catcode`\@=11
\def\@date{}
\def\figuretype{BMP}

\font\smallrm=cmr8

\font\frak=eufm10

\def\diag{\mathop{\hbox{\,\rm diag}}\nolimits}

\def\re{\mathop{\hbox{\,\rm Re}}\nolimits}

\def\tg{\mathop{\hbox{\,\rm tg}}\nolimits}

\def\spann{\mathop{\hbox{\,\rm Span}}\nolimits}
\def\hc{\hbox{\bf C}}
\def\hr{\hbox{\bf R}}

\def\gggg{\hbox{\frak g}}
\def\hhhh{\hbox{\frak h}}

\def\llll{\hbox{\frak L}}

\def\I{\sqrt{-1}\,}
\def\@{@}

\def\section{\@startsection {section}{1}{\z@}{-3.5ex plus-1ex minus
    -.2ex}{2.3ex plus.2ex}{\reset@font\large\bf}}
\def\subsection{\@startsection{subsection}{2}{\z@}{-3.25ex plus-1ex
    minus-.2ex}{1.5ex plus.2ex}{\reset@font\bf}}
\def\subsubsection{\@startsection{subsubsection}{3}{\z@}{-3.25ex plus
 -1ex minus-.2ex}{1.5ex plus.2ex}{\reset@font\small\bf}}
\def\@afterheading{\global\@nobreaktrue\everypar{\if@nobreak
   \global\@nobreakfalse\clubpenalty \@M
   \else \clubpenalty \@clubpenalty\everypar{}\fi}}

\def\FN@{\futurelet\next}
\def\DN@{\def\next@}
\def\DNii@{\def\nextii@}
\def\RIfM@{\relax\ifmmode}
\def\RIfMIfI@{\relax\ifmmode\ifinner}
\def\setboxz@h{\setbox\z@\hbox}
\def\wdz@{\wd\z@}
\def\boxz@{\box\z@}
\def\setbox@ne{\setbox\@ne}
\def\wd@ne{\wd\@ne}
\def\Invalid@#1{\def#1{\Err@{\Invalid@@\string#1}}}
\def\Invalid@@{Invalid use of }
\def\eat@#1{}
\def\Let@{\relax\iffalse{\fi\let\\=\cr\iffalse}\fi}
\def\vspace@{\def\vspace##1{\crcr\noalign{\vskip##1\relax}}}
\newif\ifinany@
\newif\ifinalign@
\newif\ifingather@
\def\strut@{\copy\strutbox@}
\newbox\strutbox@
\setbox\strutbox@\hbox{\vrule height8\p@ depth3\p@ width\z@}
\def\topaligned{\null\,\vtop\aligned@}
\def\botaligned{\null\,\vbox\aligned@}
\def\aligned{\null\,\vcenter\aligned@}
\def\aligned@{\bgroup\vspace@\Let@
 \ifinany@\else\openup\jot\fi\ialign
 \bgroup\hfil\strut@$\m@th\displaystyle{##}$&
 $\m@th\displaystyle{{}##}$\hfil\crcr}
\def\endaligned{\crcr\egroup\egroup}

\def\alignedat#1{\null\,\vcenter\bgroup\doat@{#1}\vspace@\Let@
 \ifinany@\else\openup\jot\fi\ialign\bgroup\span\preamble@@\crcr}
\newcount\atcount@
\def\doat@#1{\toks@{\hfil\strut@$\m@th
 \displaystyle{\the\hashtoks@}$&$\m@th\displaystyle
 {{}\the\hashtoks@}$\hfil}
 \atcount@#1\relax\advance\atcount@\m@ne                                    
 \loop\ifnum\atcount@>\z@\toks@=\expandafter{\the\toks@&\hfil$\m@th
 \displaystyle{\the\hashtoks@}$&$\m@th
 \displaystyle{{}\the\hashtoks@}$\hfil}\advance
  \atcount@\m@ne\repeat                                                     
 \xdef\preamble@{\the\toks@}\xdef\preamble@@{\preamble@}}

\def\gathered{\null\,\vcenter\bgroup\vspace@\Let@
 \ifinany@\else\openup\jot\fi\ialign
 \bgroup\hfil\strut@$\m@th\displaystyle{##}$\hfil\crcr}
\def\endgathered{\crcr\egroup\egroup}
\newif\iftagsleft@
\def\TagsOnLeft{\global\tagsleft@true}
\def\TagsOnRight{\global\tagsleft@false}
\TagsOnRight
\newif\ifmathtags@
\def\TagsAsMath{\global\mathtags@true}
\def\TagsAsText{\global\mathtags@false}
\TagsAsText
\def\tagform@#1{\hbox{\rm(\ignorespaces#1\unskip)}}
\def\thetag{\leavevmode\tagform@}
\def\tag#1$${\iftagsleft@\leqno\else\eqno\fi                                
 \maketag@#1\maketag@                                                       
 $$}                                                                        
\def\maketag@{\FN@\maketag@@}
\def\maketag@@{\ifx\next"\expandafter\maketag@@@\else\expandafter\maketag@@@@
 \fi}
\def\maketag@@@"#1"#2\maketag@{\hbox{\rm#1}}                                
\def\maketag@@@@#1\maketag@{\ifmathtags@\tagform@{$\m@th#1$}\else
 \tagform@{#1}\fi}
\interdisplaylinepenalty\@M
\def\allowdisplaybreaks{\RIfMIfI@
 \onlydmatherr@\allowdisplaybreaks\else
 \interdisplaylinepenalty\z@\fi\else\onlydmatherr@\allowdisplaybreaks\fi}
\Invalid@\allowdisplaybreak
\Invalid@\displaybreak
\Invalid@\intertext
\def\allowdisplaybreak@{\def\allowdisplaybreak{\crcr\noalign{\allowbreak}}}
\def\displaybreak@{\def\displaybreak{\crcr\noalign{\break}}}
\def\intertext@{\def\intertext##1{\crcr\noalign{\vskip\belowdisplayskip
 \vbox{\normalbaselines\noindent##1}\vskip\abovedisplayskip}}}
\newskip\centering@
\centering@\z@ plus\@m\p@
\def\align{\relax\ifingather@\DN@{\csname align (in
  \string\gather)\endcsname}\else
 \ifmmode\ifinner\DN@{\onlydmatherr@\align}\else
  \let\next@\align@\fi
 \else\DN@{\onlydmatherr@\align}\fi\fi\next@}
\newhelp\andhelp@
{An extra & here is so disastrous that you should probably exit^^J
and fix things up.}
\newif\iftag@
\newcount\and@
\def\align@{\inalign@true\inany@true
 \vspace@\allowdisplaybreak@\displaybreak@\intertext@
 \def\tag{\global\tag@true\ifnum\and@=\z@\DN@{&&}\else
          \DN@{&}\fi\next@}%
 \iftagsleft@\DN@{\csname align \endcsname}\else
  \DN@{\csname align \space\endcsname}\fi\next@}
\def\Tag@{\iftag@\else\errhelp\andhelp@\err@{Extra & on this line}\fi}
\newdimen\lwidth@
\newdimen\rwidth@
\newdimen\maxlwidth@
\newdimen\maxrwidth@
\newdimen\totwidth@
\def\measure@#1\endalign{\lwidth@\z@\rwidth@\z@\maxlwidth@\z@\maxrwidth@\z@
 \global\and@\z@                                                            
 \setbox@ne\vbox                                                            
  {\everycr{\noalign{\global\tag@false\global\and@\z@}}\Let@                
  \halign{\setboxz@h{$\m@th\displaystyle{\@lign##}$}
   \global\lwidth@\wdz@                                                     
   \ifdim\lwidth@>\maxlwidth@\global\maxlwidth@\lwidth@\fi                  
   \global\advance\and@\@ne                                                 
   &\setboxz@h{$\m@th\displaystyle{{}\@lign##}$}\global\rwidth@\wdz@        
   \ifdim\rwidth@>\maxrwidth@\global\maxrwidth@\rwidth@\fi                  
   \global\advance\and@\@ne                                                
   &\Tag@
   \eat@{##}\crcr#1\crcr}}
 \totwidth@\maxlwidth@\advance\totwidth@\maxrwidth@}                       
\def\displ@y@{\global\dt@ptrue\openup\jot
 \everycr{\noalign{\global\tag@false\global\and@\z@\ifdt@p\global\dt@pfalse
 \vskip-\lineskiplimit\vskip\normallineskiplimit\else
 \penalty\interdisplaylinepenalty\fi}}}
\def\black@#1{\noalign{\ifdim#1>\displaywidth
 \dimen@\prevdepth\nointerlineskip                                          
 \vskip-\ht\strutbox@\vskip-\dp\strutbox@                                   
 \vbox{\noindent\hbox to#1{\strut@\hfill}}
 \prevdepth\dimen@                                                          
 \fi}}
\expandafter\def\csname align \space\endcsname#1\endalign
 {\measure@#1\endalign\global\and@\z@                                       
 \ifingather@\everycr{\noalign{\global\and@\z@}}\else\displ@y@\fi           
 \Let@\tabskip\centering@                                                   
 \halign to\displaywidth
  {\hfil\strut@\setboxz@h{$\m@th\displaystyle{\@lign##}$}
  \global\lwidth@\wdz@\boxz@\global\advance\and@\@ne                        
  \tabskip\z@skip                                                           
  &\setboxz@h{$\m@th\displaystyle{{}\@lign##}$}
  \global\rwidth@\wdz@\boxz@\hfill\global\advance\and@\@ne                  
  \tabskip\centering@                                                       
  &\setboxz@h{\@lign\strut@\maketag@##\maketag@}
  \dimen@\displaywidth\advance\dimen@-\totwidth@
  \divide\dimen@\tw@\advance\dimen@\maxrwidth@\advance\dimen@-\rwidth@     
  \ifdim\dimen@<\tw@\wdz@\llap{\vtop{\normalbaselines\null\boxz@}}
  \else\llap{\boxz@}\fi                                                    
  \tabskip\z@skip                                                          
  \crcr#1\crcr                                                             
  \black@\totwidth@}}                                                      
\newdimen\lineht@
\expandafter\def\csname align \endcsname#1\endalign{\measure@#1\endalign
 \global\and@\z@
 \ifdim\totwidth@>\displaywidth\let\displaywidth@\totwidth@\else
  \let\displaywidth@\displaywidth\fi                                        
 \ifingather@\everycr{\noalign{\global\and@\z@}}\else\displ@y@\fi
 \Let@\tabskip\centering@\halign to\displaywidth
  {\hfil\strut@\setboxz@h{$\m@th\displaystyle{\@lign##}$}%
  \global\lwidth@\wdz@\global\lineht@\ht\z@                                 
  \boxz@\global\advance\and@\@ne
  \tabskip\z@skip&\setboxz@h{$\m@th\displaystyle{{}\@lign##}$}%
  \global\rwidth@\wdz@\ifdim\ht\z@>\lineht@\global\lineht@\ht\z@\fi         
  \boxz@\hfil\global\advance\and@\@ne
  \tabskip\centering@&\kern-\displaywidth@                                  
  \setboxz@h{\@lign\strut@\maketag@##\maketag@}%
  \dimen@\displaywidth\advance\dimen@-\totwidth@
  \divide\dimen@\tw@\advance\dimen@\maxlwidth@\advance\dimen@-\lwidth@
  \ifdim\dimen@<\tw@\wdz@
   \rlap{\vbox{\normalbaselines\boxz@\vbox to\lineht@{}}}\else
   \rlap{\boxz@}\fi
  \tabskip\displaywidth@\crcr#1\crcr\black@\totwidth@}}
\expandafter\def\csname align (in \string\gather)\endcsname
  #1\endalign{\vcenter{\align@#1\endalign}}
\Invalid@\endalign
\newif\ifxat@
\def\alignat{\RIfMIfI@\DN@{\onlydmatherr@\alignat}\else
 \DN@{\csname alignat \endcsname}\fi\else
 \DN@{\onlydmatherr@\alignat}\fi\next@}
\newif\ifmeasuring@
\newbox\savealignat@
\expandafter\def\csname alignat \endcsname#1#2\endalignat                   
 {\inany@true\xat@false
 \def\tag{\global\tag@true\count@#1\relax\multiply\count@\tw@
  \xdef\tag@{}\loop\ifnum\count@>\and@\xdef\tag@{&\tag@}\advance\count@\m@ne
  \repeat\tag@}%
 \vspace@\allowdisplaybreak@\displaybreak@\intertext@
 \displ@y@\measuring@true                                                   
 \setbox\savealignat@\hbox{$\m@th\displaystyle\Let@
  \attag@{#1}
  \vbox{\halign{\span\preamble@@\crcr#2\crcr}}$}%
 \measuring@false                                                           
 \Let@\attag@{#1}
 \tabskip\centering@\halign to\displaywidth
  {\span\preamble@@\crcr#2\crcr                                             
  \black@{\wd\savealignat@}}}                                               
\Invalid@\endalignat

\def\matrix{\,\vcenter\bgroup\Let@\vspace@
    \normalbaselines
  \m@th\ialign\bgroup\hfil$##$\hfil&&\quad\hfil$##$\hfil\crcr
    \mathstrut\crcr\noalign{\kern-\baselineskip}}
\def\endmatrix{\crcr\mathstrut\crcr\noalign{\kern-\baselineskip}\egroup
                \egroup\,}
\def\pmatrix{\left(\matrix}
\def\endpmatrix{\endmatrix\right)}

\newdimen\spreadmlines@
\def\format{\crcr\egroup\iffalse{\fi\ifnum`}=0 \fi\format@}
\newtoks\hashtoks@
\hashtoks@{#}
\def\format@#1\\{\def\preamble@{#1}%
 \def\l{$\m@th\the\hashtoks@$\hfil}%
 \def\c{\hfil$\m@th\the\hashtoks@$\hfil}%
 \def\r{\hfil$\m@th\the\hashtoks@$}%
 \edef\Preamble@{\preamble@}\ifnum`{=0 \fi\iffalse}\fi
 \ialign\bgroup\span\Preamble@\crcr}
\def\cases{\bgroup\spreadmlines@\jot\left\{\,\matrix\format\l&\quad\l\\}
\def\endcases{\endmatrix\right.\egroup}

\def\wd@ne{\wd\@ne}
\def\setbox@ne{\setbox\@ne}
\def\binrel@#1{\setboxz@h{\thinmuskip0mu
  \medmuskip\m@ne mu\thickmuskip\@ne mu$#1\m@th$}%
 \setbox@ne\hbox{\thinmuskip0mu\medmuskip\m@ne mu\thickmuskip
  \@ne mu${}#1{}\m@th$}%
 \setbox\tw@\hbox{\hskip\wd@ne\hskip-\wdz@}}
\def\overset#1\to#2{\binrel@{#2}\ifdim\wd\tw@<\z@
 \mathbin{\mathop{\kern\z@#2}\limits^{#1}}\else\ifdim\wd\tw@>\z@
 \mathrel{\mathop{\kern\z@#2}\limits^{#1}}\else
 {\mathop{\kern\z@#2}\limits^{#1}}{}\fi\fi}
\def\underset#1\to#2{\binrel@{#2}\ifdim\wd\tw@<\z@
 \mathbin{\mathop{\kern\z@#2}\limits_{#1}}\else\ifdim\wd\tw@>\z@
 \mathrel{\mathop{\kern\z@#2}\limits_{#1}}\else
 {\mathop{\kern\z@#2}\limits_{#1}}{}\fi\fi}

\def\setboxz@h{\setbox\z@\hbox}
\def\wdz@{\wd\z@}
\def\boxz@{\box\z@}
\newif\ifmsbmloaded@
\def\widetilde#1{
\ifmsbmloaded@
  \setboxz@h{$\m@th#1$}\ifdim\wdz@>\tw@ em\mathaccent"0\msbfam@5D{#1}\else
  \mathaccent"0365{#1}\fi
 \else\mathaccent"0365{#1}\fi}

\mathchardef\varGamma="0100
\mathchardef\varDelta="0101
\mathchardef\varTheta="0102
\mathchardef\varLambda="0103
\mathchardef\varXi="0104
\mathchardef\varPi="0105
\mathchardef\varSigma="0106
\mathchardef\varUpsilon="0107
\mathchardef\varPhi="0108
\mathchardef\varPsi="0109
\mathchardef\varOmega="010A

\def\allowmathbreak{\relax\ifmmode\ifinner\allowbreak\else
  \nonmathaerr@\allowmathbreak\fi\else\nonmathberr@\allowmathbreak\fi}

\def\theoremfont{\def\@thmfont{\rm}}
\@addtoreset{definition}{section}

\newtheorem{theorem}{\bf Theorem}
\newtheorem{lemma}{\bf Lemma}

\newtheorem{remark}{\sl Remark}
\def\demo{\vskip2pt\noindent{\sl Proof.}\hskip6pt\relax}
\def\enddemo{\vskip2pt\relax}

\def\dsize{\displaystyle}
\def\D{\dsize}

\def\bigskip{\vskip12pt}
\def\medskip{\vskip6pt}
\def\smallskip{\vskip2pt}

\def\voidtoken{}

\def\refby{}
\def\refpaper{}
\def\refjour{}
\def\refvol{}
\def\refpage{}
\def\refpages{}
\def\refyr{}
\def\refbook{}
\def\refpubl{}
\def\refpubladdr{}

\def\bibref{
  \global\def\refby{}
  \global\def\refpaper{}
  \global\def\refjour{}
  \global\def\refvol{}
  \global\def\refpage{}
  \global\def\refpages{}
  \global\def\refyr{}
  \global\def\refbook{}
  \global\def\refpubl{}
  \global\def\refpubladdr{}
  }
\def\by#1{\global\def\refby{#1}}
\def\paper#1{\global\def\refpaper{#1}}
\def\jour#1{\global\def\refjour{#1}}
\def\vol#1{\global\def\refvol{#1}}
\def\page#1{\global\def\refpage{#1}}
\def\pages#1{\global\def\refpage{#1}}
\def\yr#1{\global\def\refyr{#1}}
\def\book#1{\global\def\refbook{#1}}
\def\publ#1{\global\def\refpubl{#1}}
\def\publaddr#1{\global\def\refpubladdr{#1}}
\def\endbibref{{
\ifx\refby\voidtoken \else \refby\fi
\ifx\refpaper\voidtoken \else , {\it\refpaper\/}\fi
\ifx\refjour\voidtoken \else , \refjour\fi
\ifx\refbook\voidtoken \else , \refbook\fi
\ifx\refpubl\voidtoken \else , \refpubl\fi
\ifx\refpubladdr\voidtoken \else , \refpubladdr\fi
\ifx\refvol\voidtoken \else \ \refvol\fi
\ifx\refyr\voidtoken \else \ (\refyr)\fi
\ifx\refpage\voidtoken \else , \refpage\fi.\vskip2pt}}

\newif\iff@rstzcite\f@rstzcitetrue
\def\zcite{\@tempswafalse\@zcitex[]}
\def\@zcitex[#1]#2{\if@filesw\immediate\write\@auxout{\string\citation{#2}}\fi
  \let\@citea\@empty
  \@zcite{\@for\@citeb:=#2\do
    {\@citea\def\@citea{\raise.75ex\hbox{\scriptsize,}\penalty\@m}%
     \def\@tempa##1##2\@nil{\edef\@citeb{\if##1\space##2\else##1##2\fi}}%
     \expandafter\@tempa\@citeb\@nil
     \@ifundefined{b@\@citeb}{{\reset@font\bf ? }\@warning
       {Citation `\@citeb' on page \thepage \space undefined}}
     \hbox{{}\raise0.75ex\hbox{\scriptsize\iff@rstzcite [\hskip-0.05em \fi
     \csname b@\@citeb\endcsname}}\f@rstzcitefalse}}{#1}
     \hskip-0.3em\hbox{{}\raise0.75ex\hbox{\scriptsize]}}\f@rstzcitetrue}
\def\@zcite#1{#1} 

\def\maketitle{\par
 \begingroup
   \if@twocolumn
     \twocolumn[\@maketitle]%
     \else 
     \global\@topnum\z@
     \@maketitle \fi
     \@thanks
 \endgroup
 \setcounter{footnote}{0}%
 \let\maketitle\relax
 \let\@maketitle\relax
 \gdef\@thanks{}\gdef\@author{}\gdef\@title{}\let\thanks\relax}

\def\@maketitle{
 \null
 \vskip 2em
 \begin{center}%
  {\LARGE \@title \par}%
  \vskip 1.5em
  {\large
   \lineskip .5em
   \begin{tabular}[t]{c}\@author
   \end{tabular}\par}%
  \vskip 1em
  {\large \@date}%
 \end{center}%
 \par
 \vskip 1.5em}

\def\BMPtype{BMP}

\catcode`\@=\active

\def\figuretype{EPS}

\def\soex{so_{\hbox{\smallrm ex}}}

\begin{document}

\title{Darboux transformations for twisted so(p,q) system
and local isometric immersion of space forms}

\author{\small Zixiang Zhou\\
\small Institute of Mathematics, Fudan University,Shanghai 200433, China\\
\small E-mail: zxzhou\@guomai.sh.cn
}

\maketitle

\abstract
For the $n$-dimensional integrable system with a twisted $so(p,q)$
reduction, Darboux transformations given by Darboux matrices of
degree $2$ are constructed explicitly. These Darboux
transformations are applied to the local isometric immersion of 
space forms with flat normal bundle and linearly independent
curvature normals to give the explicit expression of the position
vector. Some examples are given from the trivial solutions and
standard imbedding $T^n\to\hr^{2n}$. 
\endabstract

\section{Introduction}

The theory of integrable system has been widely used to study
some differential geometric problems such as minimal
submanifolds, submanifolds with constant mean curvature, harmonic
maps etc. Especially, the isometric immersion of space form
$M_1(K_1)$ of curvature $K_1$ into space form $M_2(K_2)$ of
curvature $K_2$ was studied in various papers. If $K_1\ne K_2$,
the nonlinear wave equation and the nonlinear sine-Gordon
equation are considered to describe some special
problems.\zcite{bib:ABT,bib:BFT,bib:BT,bib:GH,bib:Tenen} When 
$K_1=K_2$, the local isometric immersion with flat normal bundle
and linearly independent curvature normals was proposed in
\cite{bib:Terng}. That problem was also dealt with by a purely
geometric way.\zcite{bib:DajCan,bib:DajReine}

In this paper, we use the Darboux transformation to get the
explicit expressions of the local isometric immersions of the
space forms of the same curvature with flat normal bundle
and linearly independent curvature normals. 

In this problem, the Lax pair has a twisted $so(n)$ reduction. 
When the Lie algebra is $gl(n,\hc)$ or $sl(n,\hc)$, there is a
systematic construction of Darboux transformations (for the
problems discussed later, see
\cite{bib:GuNankai,bib:GuNdim,bib:MS,bib:SZ}), which is now a
useful method to get explicit solutions of nonlinear integrable
partial differential equations. If the Lie algebra is $su(p,q)$,
there is also a general algorithm to choose the spectral
parameters\zcite{bib:GZNdim,bib:MS,bib:NMPZ,bib:Zhou2n} for the
Darboux transformation of degree $1$. In this algorithm, the
spectral parameters can take only two mutually conjugate values.
As a subalgebra of $su(p,q)$ ($p+q$ even), $so(p,q)$ ($p+q$ even)
problem can be dealt with in a similar way, provided that the
real condition can be realized. However, for $so(p,q)$ ($p+q$
odd) problem, this method is not applicable
directly\zcite{bib:Cie,bib:GZpcf} since the spectral parameters
should be conjugate and cannot be real (or purely imaginary if
written in another way) so that the Darboux transformation is not
trivial. It is known that the pure $so(p,q)$ problem can be dealt
with by a Darboux transformation of degree
$2$.\zcite{bib:Gerd,bib:ZM} Here, for the twisted $so(p,q)$ problem
($so(p,q)$ problem with an additional involution condition), we
construct such a Darboux transformation using a limit process so
that the Darboux matrix has only two (not four) eigenvalues. For
the twisted $so(p,q)$ system, the method in \cite{bib:ZM} can
give the same Darboux matrix under some assumptions, but the
method here is more direct and has no assumptions. 

\S\ref{sec:ls} describes the linear system and the properties of
its solutions. In \S\ref{sec:dt}, we get explicit expressions of
the Darboux transformations of degree $2$ for twisted $so(p,q)$ 
reduction. Owing to the isomorphism between $su(1,1)$ and
$so(2,1)$, we can see that the Darboux transformation for MKdV
and sine-Gordon equations given by the Darboux matrix here is
actually the well-known standard Darboux transformation. 

Using the above conclusions, \S\ref{sec:geo} gives the Darboux
transformation for the local isometric immersion from $M_n(K)$ to
$M_{2n}(K)$ with flat normal bundle and linearly independent
curvature normals, the Lax set of which was proposed by
\cite{bib:Terng}. We present the general expression of the
transformation for the position vector of $M_n(K)\to M_{2n}(K)$.
\S\ref{sec:geo} also gives some interesting examples, including
the submanifolds derived from trivial solutions for all
$K=0,1,-1$ cases and the submanifold derived from the standard
torus $T^n$ in $\hr^{2n}$.  

\section{Linear system}\label{sec:ls}

Let 
\begin{equation}
   \gggg=\soex(p,q,r)=\{\,X\in gl(p+q+r,\hr)\,|\,X^TC+CX=0\,\}
\end{equation}
where $C=I_{p,q,r}=\diag(\underbrace{1,\cdots,1}_p,
\underbrace{-1,\cdots,-1}_q,\underbrace{0,\cdots,0}_r)$.  
Clearly, $\soex(p,q,0)=so(p,q)$. Here we consider $\soex(p,q,r)$
instead of $so(p,q)$ for the unified treatment in \S\ref{sec:geo}.

Let $\sigma$ be a diagonal matrix such that $\sigma^2=1$
($\sigma\ne 1$). Then the transformation $\theta:\gggg\to\gggg$,
$X\mapsto\sigma X\sigma$ is an involution on $\gggg$. Hence there
is a decomposition $\gggg=\gggg_0\oplus\gggg_1$ where $\gggg_0$
and $\gggg_1$ are the $+1$ and $-1$ eigenspaces of $\theta$
respectively. Moreover, we suppose that there is a maximum
commutative subalgebra $\hhhh$ of $\gggg$ in $\gggg_1$. 

Let 
\begin{equation}
 \llll(\gggg)=\big\{\,\sum_{j=0}^n
   X_j\lambda^j\,\big|\,X_j\in\gggg,j=0,1,\cdots,n\,\big\}
\end{equation}
be a subalgebra of the loop algebra of $\gggg$,
\begin{equation}
 \llll^\sigma(\gggg)=\{\,v\in\llll(\gggg)\,|
   \,\sigma v(\lambda)\sigma=v(-\lambda)\,\} 
\end{equation} 
be a subalgebra of $\llll(\gggg)$.

First we consider the linear system
\begin{equation}
   \varPhi_x=U(x,\lambda)\varPhi \label{eq:lp}
\end{equation}
where $U\in\llll^\sigma(\gggg)$. Denote
\begin{equation}
   U(x,\lambda)=\sum_{j=0}^n U_j(x)\lambda^j,\quad 
   U_j(x)\in\gggg_{[j]}, \quad 
   [j]=\cases 0&j\hbox{ even,}\\ 1&j\hbox{ odd,} \endcases
\end{equation}
then $U\in\llll^\sigma(\gggg)$ if and only if $\sigma
U(x,\lambda)\sigma=U(x,-\lambda)$,
$(U(x,\lambda))^*C+CU(x,\bar\lambda)=0$. In this paper, we always
suppose $\varPhi(x,\lambda)$, a solution of (\ref{eq:lp}), is smooth
with respect to both $x$ and $\lambda$. To discuss the Darboux 
transformation for (\ref{eq:lp}), we need the following lemmas.
 
\begin{lemma}\label{lemma1}
If $\varPhi$ is a solution of (\ref{eq:lp}) with $\lambda=\lambda_0$,
then $\sigma\varPhi$ is a solution of (\ref{eq:lp}) with
$\lambda=-\lambda_0$. 
\end{lemma}

\begin{lemma}\label{lemma2}
If $\varPhi$ is a solution of (\ref{eq:lp}) with
$\lambda=\lambda_0$, $\varPsi$ is a solution of (\ref{eq:lp})
with $\lambda=\bar\lambda_0$, then $(\varPsi^*C\varPhi)_x=0$. 
\end{lemma}

\demo
The conclusion follows from
\begin{equation}
 \aligned
   &\varPhi_x=U(x,\lambda_0)\varPhi,\\
   &\varPsi_x^*C=-\varPsi^*CU(x,\lambda_0).
   \endaligned
\end{equation}
\enddemo

\begin{lemma}\label{lemma3}
Suppose $\mu\in\hr$, $\varPhi$ is a solution of (\ref{eq:lp})
such that $\varPhi\big|_{\lambda=\mu}$ is real, then
$\varPhi^TC\varPhi_\lambda\big|_{\lambda=\mu}$ is independent of
$x$. 
\end{lemma}

\demo
\begin{equation}
 \aligned
   &(\varPhi_\lambda)_x=U(x,\lambda)\varPhi_\lambda
    +U_\lambda(x,\lambda)\varPhi, \\
   &\varPhi_x^*C=-\varPhi^*CU(x,\lambda),
   \endaligned
\end{equation}
hold at $\lambda=\mu$, hence
\begin{equation}
 (\varPhi^*C\varPhi_\lambda)_x\big|_{\lambda=\mu}
   =\varPhi^*CU_\lambda(x,\lambda)\varPhi\big|_{\lambda=\mu}=0
\end{equation}
since $CX$ is antisymmetric for any $X\in\gggg$.
\enddemo

\section{Darboux transformations}\label{sec:dt}

We say a real symmetric matrix $M$ is semidefinite if for any
nonzero vector $\xi$, $\xi^TM\xi\ge 0$ or $\xi^TM\xi\le 0$. The
Darboux transformation for (\ref{eq:lp}) can be constructed in 
the following two ways. The first one is applicable when $C$ is
not semidefinite and the second one is applicable when $C\sigma$
is not semidefinite. Since $\sigma$ is not $\pm I$, these two
cases cover all the possible situations. 

\subsection{Construction of Darboux transformation when $C$
is not semidefinite}\label{subsec:dt1}

Let $\mu\in\hr$. Let $H$ be a real vector solution of (\ref{eq:lp})
with $\lambda=\mu$ and satisfies $H^*CH=0$. (This is possible due
to Lemma~\ref{lemma2}.) 

Take $\lambda_1^{(\varepsilon)}=\mu+\I\varepsilon$,
$\lambda_2^{(\varepsilon)}=-\mu-\I\varepsilon$. Let
$h_1^{(\varepsilon)}$ be a vector solution of (\ref{eq:lp}) with
$\lambda=\lambda_1^{(\varepsilon)}$ satisfying $h_1^{(0)}=H$ and
$h_1^{(\varepsilon)}|_{x=x_0}=H(x_0)$ for some fixed $x_0$.
According to Lemma~\ref{lemma1}, $h_2^{(\varepsilon)}=\sigma
h_1^{(\varepsilon)}$ is a solution of (\ref{eq:lp}) with
$\lambda=\lambda_2^{(\varepsilon)}$. 

Let
\begin{equation}
 \varGamma_{jk}^{(\varepsilon)}
    =\frac{h_j^{(\varepsilon)*}Ch_k^{(\varepsilon)}}
    {\lambda_k^{(\varepsilon)}-\bar\lambda_j^{(\varepsilon)}}
   \qquad (j,k=1,2),
\end{equation}
then
\begin{equation}
 G^{(\varepsilon)}(\lambda)
   =1-\sum_{j,k=1}^2
   \frac 1{\lambda-\bar\lambda_k^{(\varepsilon)}}
   h_j^{(\varepsilon)}(\varGamma^{(\varepsilon)-1})_{jk}
   h_k^{(\varepsilon)*}C
\end{equation}
is a Darboux matrix for (\ref{eq:lp}) without considering the
$L^\sigma(\gggg)$ reduction, that is, for any solution $\varPhi$
of (\ref{eq:lp}), $\widetilde\varPhi=G^{(\varepsilon)}\varPhi$
satisfies 
\begin{equation}
   \widetilde\varPhi_x=\widetilde U^{(\varepsilon)}(x,\lambda)
   \widetilde\varPhi
\end{equation}
where $\widetilde
U^{(\varepsilon)}\in\llll(gl(p+q+r,\hc))$\zcite{bib:Zhou2n}. Now 
we calculate $\varGamma^{(\varepsilon)}$, $G^{(\varepsilon)}$ and
their limits as $\varepsilon\to 0$. First, 
\begin{equation}
   \varGamma^{(\varepsilon)}=\pmatrix
   \dsize\frac{h_1^{(\varepsilon)*}Ch_1^{(\varepsilon)}}{2\I\varepsilon}
    &\dsize\frac{h_1^{(\varepsilon)*}C\sigma h_1^{(\varepsilon)}}{-2\mu} \\
   \dsize\frac{h_1^{(\varepsilon)*}C\sigma h_1^{(\varepsilon)}}{2\mu}
    &\dsize\frac{h_1^{(\varepsilon)*}Ch_1^{(\varepsilon)}}{-2\I\varepsilon}
   \endpmatrix.
\end{equation}
By Lemma~\ref{lemma3} and the assumptions $H^*CH=0$,
$\dsize\left.\frac{\partial
h_1^{(\varepsilon)}}{\partial\varepsilon}\right|_{x=x_0}=0$, 
\begin{equation}
 \aligned
   &\lim_{\varepsilon\to 0}\frac{h_1^{(\varepsilon)*}C
    h_1^{(\varepsilon)}}{2\I\varepsilon}
   =-\I\lim_{\varepsilon\to 0}
    \re\bigg(h_1^{(\varepsilon)*}C\frac{\partial h_1^{(\varepsilon)}}
    {\partial\varepsilon}\bigg) \\
   &=-\I \lim_{\varepsilon\to 0}
    \re\bigg(h_1^{(\varepsilon)*}C\frac{\partial h_1^{(\varepsilon)}}
    {\partial\varepsilon}\bigg)\bigg|_{x=x_0}=0. 
   \endaligned
\end{equation}
Hence
\begin{equation}
 \varGamma=\lim_{\varepsilon\to 0}\varGamma^{(\varepsilon)}
    =\frac 1{2\mu}
    \pmatrix 0 &-H^TC\sigma H\\ H^TC\sigma H &0\endpmatrix
    =\frac{H^TC\sigma H}{2\mu}\pmatrix 0&-1\\ 1&0\endpmatrix.
\end{equation}
\begin{equation}
 \varGamma^{-1}=\frac{2\mu}{H^TC\sigma H} 
    \pmatrix 0 &1\\ -1 &0\endpmatrix.
\end{equation}
\begin{equation}
 \aligned
   G(\lambda)&=\lim_{\varepsilon\to 0}G^{(\varepsilon)}(\lambda)
    =1-\frac{2\mu}{H^TC\sigma H}
    \left(\frac{HH^T\sigma}{\lambda+\mu}
    -\frac{\sigma HH^T}{\lambda-\mu}\right)C \\
   &=\frac 1{\lambda^2-\mu^2}\left(\lambda^2
    +\frac{2\lambda\mu}{H^TC\sigma H}[\sigma,HH^T]C
    +\frac{2\mu^2(\sigma HH^T+HH^T\sigma)C}{H^TC\sigma H}-\mu^2\right).
   \endaligned
\end{equation}
It is easy to prove that
\begin{equation}
   \aligned
   &(G(\bar\lambda))^*CG(\lambda)=C, \\
   &\sigma G(\lambda)\sigma=G(-\lambda). 
   \endaligned 
\end{equation}
Using these facts, we know that
\begin{equation}
 \widetilde U=GUG^{-1}+G_xG^{-1}
\end{equation}
satisfies $\widetilde U\in\llll^\sigma(\gggg)$.

Therefore, the following theorem holds.

\begin{theorem}\label{thm1}
Suppose $C$ is not semidefinite. Let $\mu\in\hr$. Let $H$ be a
real vector solution of (\ref{eq:lp}) with $\lambda=\mu$ such
that $H^TCH=0$. Let
\begin{equation}
   \aligned
   &G(\lambda)=\frac 1{\lambda^2-\mu^2}\left(\lambda^2
    +\frac{2\lambda\mu}{H^TC\sigma H}[\sigma,HH^T]C
    +\frac{2\mu^2(\sigma HH^T+HH^T\sigma)C}
    {H^TC\sigma H}-\mu^2\right), \\
   &\widetilde U=GUG^{-1}+G_xG^{-1},
   \endaligned 
\end{equation}
then $\widetilde U\in\llll^\sigma(\gggg)$. Moreover, for any
solution $\varPhi$ of (\ref{eq:lp}), $\widetilde\varPhi=G\varPhi$
satisfies $\widetilde\varPhi_x=\widetilde U\widetilde\varPhi$. 
\end{theorem}

\subsection{Construction of Darboux transformation when $C\sigma$
is not semidefinite}\label{subsec:dt2}

When $C$ is semidefinite, Theorem~\ref{thm1} is no longer valid,
since the required $H$ does not exist in general. Here we discuss
the problem when $C\sigma$ is not semidefinite. We use the 
following transformation to change the problem to a problem dealt
with in Part \ref{subsec:dt1}.

Consider the linear system
\begin{equation}
   \varPhi_x=U(\I\lambda)\varPhi.
   \label{eq:lp2}
\end{equation}
Let $\tau$ be a
complex diagonal matrix such that $\tau^2=\sigma$ and $\tau$ has
only two eigenvalues $1$, $\I$. Clearly $\tau^*\tau=1$. Let
$V(\lambda)=\tau U(\I\lambda)\tau^*$, then when $\lambda$ is real,
\begin{equation}
 \aligned
   &\hbox{(1)}\quad
    \overline{V(\lambda)}=\tau^*U(-\I\lambda)\tau
    =\tau^*\sigma U(\I\lambda)\sigma\tau
    =\tau U(\I\lambda)\tau^*=V(\lambda),\\
   &\hbox{(2)}\quad
    (V(\lambda))^*C\sigma=\tau(U(\I\lambda))^*\tau^*C\sigma
    =-\tau CU(-\I\lambda)\tau \\
   &\qquad=-\tau C\sigma U(\I\lambda)\tau^*
    =-\sigma CV(\lambda), \\
   &\hbox{(3)}\quad
    V(-\lambda)=\tau U(-\I\lambda)\tau^*
    =\tau\sigma U(\I\lambda)\sigma\tau^*=\sigma V(\lambda)\sigma.
\endaligned
\end{equation}
Hence, $V(\lambda)\in L^\sigma(\gggg')$ where
\begin{equation}
   \gggg'=\{\,X\in gl(p+q+r,\hr)\,|\,
   X^TC\sigma+C\sigma X=0\,\}\cong \soex(p',q',r)
\end{equation}
with $p'+q'=p+q$.

Let $\varPsi=\tau\varPhi$, then the linear system (\ref{eq:lp2})
is changed to 
\begin{equation}
   \varPsi_x=V(\lambda)\varPsi.
   \label{eq:lpV}
\end{equation}
When $C\sigma$ is not semidefinite, we can use Theorem~\ref{thm1}
to this problem.

Let $G'(\lambda)=\tau G(\I\lambda)\tau^*$, then
$\widetilde\varPsi=G'(\lambda)\varPsi$ satisfies 
$$
   \widetilde\varPsi_x=\widetilde V(\lambda)\varPsi
$$
where $\widetilde V(\lambda)=\tau \widetilde U(\I\lambda)\tau^*$.

Let $\mu\in\hr$. Let $H'$ be a real solution of (\ref{eq:lpV})
with $\lambda=-\mu$. Then Theorem~\ref{thm1} implies that
$G'(\lambda)$ can be chosen as 
$$ 
   G'(\lambda')=\frac 1{\lambda^{\prime2}-\mu^2}\left(\lambda^{\prime2}
    -\frac{2\lambda'\mu}{H^{\prime T}C\sigma H'}[\sigma,H'H^{\prime T}]
    \sigma C
    +\frac{2\mu^2(\sigma H'H^{\prime T}+H'H^{\prime T}\sigma)\sigma C}
    {H^{\prime T}C\sigma H'}-\mu^2\right)
$$
since $C$ should be replaced by $\sigma C$ here. However,
$\tau^{-1} H'$ is a solution of (\ref{eq:lp2}) with
$\lambda=-\mu$, i.e. it is a solution of (\ref{eq:lp})
with $\lambda=\I\mu$. Let $H=\tau^{-1} H'$, then, for
$\lambda=-\I\lambda'$, 
$$ \aligned
   &G(\lambda)=\tau^{-1} G(\lambda')\tau\\
   &=\frac 1{\lambda^2+\mu^2}\left(\lambda^2
    +\frac{2\I\lambda\mu}{H^*CH}[\sigma,HH^*]\sigma C
    -\frac{2\mu^2(\sigma HH^*+HH^*\sigma)\sigma C}{H^*CH}+\mu^2\right).
   \endaligned $$
Therefore, we have

\begin{theorem}\label{thm2}
Suppose $C\sigma$ is not semidefinite. Let $\mu\in\hr$. Let $H$
be a complex vector solution of (\ref{eq:lp}) with $\lambda=\I\mu$
such that $\tau H$ is real and $H^*C\sigma H=0$. Let
\begin{equation}
 \aligned
   &G(\lambda)=\frac 1{\lambda^2+\mu^2}\left(\lambda^2
    +\frac{2\I\lambda\mu}{H^*CH}[\sigma,HH^*]\sigma C
    -\frac{2\mu^2(\sigma HH^*+HH^*\sigma)\sigma C}{H^*CH}+\mu^2\right), \\
   &\widetilde U=GUG^{-1}+G_xG^{-1},
   \endaligned
\end{equation}
then $\widetilde U\in\llll^\sigma(\gggg)$. Moreover, for any
solution $\varPhi$ of (\ref{eq:lp}), $\widetilde\varPhi=G\varPhi$
satisfies $\widetilde\varPhi_x=\widetilde U\widetilde\varPhi$. 
\end{theorem}

\subsection{Example: MKdV equation and sine-Gordon equation}

The MKdV equation
\begin{equation}
 u_t=u_{xxx}+6u^2u_x
\end{equation}
has a well-known Lax pair
\begin{equation}
 \aligned
   &\varPsi_x=\frac\lambda 2\pmatrix 1&0\\ 0&-1\endpmatrix\varPsi
    +\pmatrix 0&u\\ -u&0\endpmatrix\varPsi,\\
   &\varPsi_t=\bigg(\frac{\lambda^3}2+\lambda u^2\bigg)
    \pmatrix 1&0\\ 0&-1\endpmatrix\varPsi
    +(\lambda^2u+u_{xx}+2u^2)\pmatrix 0&u\\ -u&0\endpmatrix\varPsi
    +\lambda u_x\pmatrix 0 &1\\ 1&0\endpmatrix\varPsi.
   \endaligned
\end{equation}
The real Lie algebra $su(1,1)$ generated by
\begin{equation}
 e_1=\frac 12\pmatrix 1&0\\ 0&-1\endpmatrix,\quad
   e_2=\frac 12\pmatrix 0 &1\\ 1&0\endpmatrix,\quad
   e_3=\frac 12\pmatrix 0&1\\ -1&0\endpmatrix
\end{equation}
is isomorphic to $so(2,1)$, given by the correspondence
\begin{equation}
 e_1\to\pmatrix 0&0&0\\ 0&0&1\\ 0&1&0\endpmatrix,\quad
   e_2\to\pmatrix 0&0&-1\\ 0&0&0\\ -1&0&0\endpmatrix,\quad
   e_3\to\pmatrix 0&1&0\\ -1&0&0\\ 0&0&0\endpmatrix. 
   \label{eq:su2so3}
\end{equation}
Now take $\gggg=so(2,1)$, $\dsize C=\pmatrix 1&0&0\\ 0&1&0\\
0&0&-1\endpmatrix $, $\sigma=C$. Then,
\begin{equation}
 \aligned
   &\gggg_0=\spann\left\{
    \pmatrix 0&1&0\\ -1&0&0\\ 0&0&0\endpmatrix\right\},\\
   &\gggg_1=\spann\left\{\pmatrix 0&0&0\\ 0&0&1\\ 0&1&0\endpmatrix,\,
    \pmatrix 0&0&-1\\ 0&0&0\\ -1&0&0\endpmatrix\right\}.
   \endaligned
\end{equation}
Take the Cartan subalgebra
\begin{equation}
 \hhhh=\spann\left\{\pmatrix 0&0&0\\ 0&0&1\\
    0&1&0\endpmatrix\right\}
\end{equation}
which corresponds to $e_1$.

Using the correspondence (\ref{eq:su2so3}), we have the new Lax pair
\begin{equation}
 \aligned
   &\varPhi_x=\pmatrix 
    0  &2u     &0\\
    -2u &0       &\lambda \\
    0  &\lambda &0 
    \endpmatrix\varPhi,\\
   &\varPhi_t=\pmatrix 
    0                  &2\lambda^2u+2u_{xx}+4u^3      &-2\lambda u_x\\
    -2\lambda^2u-2u_{xx}-4u^3   &0           &\lambda^3+2\lambda u^2\\
    -2\lambda u_x      &\lambda^3+2\lambda u^2        &0
    \endpmatrix\varPhi.
   \endaligned
   \label{eq:lpMKdV3}
\end{equation}
Let $\mu\in\hr$. Let $H=(\alpha,\beta,\gamma)^T$ be a real
solution of (\ref{eq:lpMKdV3}) with $\lambda=\mu$ and satisfies
$\alpha^2+\beta^2-\gamma^2=0$, then Theorem~\ref{thm1} gives
\begin{equation}
   (\lambda^2-\mu^2)G(\lambda)=\lambda^2-\frac{2\lambda\mu}{\gamma^2}
    \pmatrix 
     0            &0           &\alpha\gamma  \\ 
     0            &0           &\beta\gamma   \\ 
     \alpha\gamma &\beta\gamma &0             \endpmatrix
    +\frac{2\mu^2}{\gamma^2}
    \pmatrix 
     \alpha^2    &\alpha\beta &0         \\ 
     \alpha\beta &\beta^2     &0         \\ 
     0           &0           &\gamma^2  \endpmatrix
    -\mu^2,
\end{equation}
\begin{equation}
 \widetilde u=u-\frac{\mu\alpha}{\gamma}. \label{eq:dtMKdV}
\end{equation}
Let $\theta=\sin^{-1}(\alpha/\gamma)$, then (\ref{eq:lpMKdV3})
gives $\theta_x=2u-\mu\sin\theta$. Therefore, (\ref{eq:dtMKdV})
is just the same as the standard Darboux transformation for MKdV
equation given by a $2\times 2$ Darboux matrix, which is linear
in the spectral parameter, although the $3\times 3$ Darboux
matrix here is quadratic in $\lambda$.  

For the sine-Gordon equation, the $x$-part of the Lax pair is the
same as that of the MKdV equation. Hence the Darboux
transformation (\ref{eq:dtMKdV}) is also the standard Darboux 
transformation for sine-Gordon equation. 

\section{Local isometric immersion of space forms with flat normal
bundle and linearly independent curvature normals}\label{sec:geo}

\cite{bib:Terng} gives the Lax sets for the equations describing
local isometric immersions with flat normal bundles and linearly
independent curvature normals. Here we apply the above theorems
to this problem and give the Darboux transformations of the
position vectors.

First we list some useful conclusions in \cite{bib:Terng},
and rewrite the structure equation to a form to which the Darboux
transformation can be easily applied. 

Let $M_n(K)$ be an $n$-dimensional space form of curvature $K$
where $K=0,1,-1$. For the local isometric immersion $M_n(K)\to
M_{2n}(K)$ with flat normal bundle and linearly independent
curvature normals, there always exist local coordinates 
$x=(x_1,\cdots,x_n)$ on $M_n(K)$ and the parallel orthonormal
normal vector fields $(e_{n+1},\cdots,e_{2n})$ such that the
first and second fundamental forms are
\begin{equation}
 \aligned
   &I=\sum_{i,j=1}^n g_{ij}\,dx_i\,dx_j
    =\sum_{i=1}^n\rho_i^2(x)\,dx_i^2,\\
   &I\!I=\sum_{i,j,\alpha=1}^n\varOmega_{ij}^\alpha\,
    dx_i\,dx_j\,e_{n+\alpha}
    =\sum_{i,j=1}^n\rho_i(x)\omega_{i\alpha}(x)\,dx_i^2\,e_{n+\alpha}
   \endaligned \label{eq:ff12}
\end{equation}
where $\omega(x)=(\omega_{ij}(x))\in O(n)$.

Denote 
\begin{equation}
 R_n(K)=\cases \hr^n, \hbox{ if } K=0,1,\\ 
   \hr^{n-1,1}, \hbox{ if } K=-1, \endcases
\end{equation}
where $\hr^{n-1,1}$ has the metric $x_1^2+\cdots+x_{n-1}^2-x_n^2$.

Let $i_n(K):M_n(K)\to R_{n+1}(K)$ be the standard imbedding given by
\begin{equation}
 \aligned
   &M_n(0)=\{\,(x_1,\cdots,x_{n+1})\in\hr^{n+1}\,|\,x_{n+1}=0\,\}, \\
   &M_n(1)=\{\,(x_1,\cdots,x_{n+1})\in\hr^{n+1}\,|\,
    x_1^2+\cdots+x_n^2+x_{n+1}^2=1\,\}, \\
   &M_n(-1)=\{\,(x_1,\cdots,x_{n+1})\in\hr^{n,1}\,|\,
    x_1^2+\cdots+x_n^2-x_{n+1}^2=-1\,\}.
   \endaligned
\end{equation}
Then we can consider the immersion
\begin{equation}
 \vec V:M_n(K)\to M_{2n}(K)\overset{i_{2n}(K)}\to\longrightarrow
   R_{2n+1}(K).
\end{equation}
Clearly, $\vec V\in\hr^{2n}$ for $K=0$, $\vec V\cdot \vec V=K$
for $K=\pm 1$. Here ``$\cdot$'' refers to the inner product in
$R_{2n+1}(K)$. 

\begin{remark} 
We imbed $\hr^{2n}$ into $\hr^{2n+1}$ only for the unification of
the three cases.
\end{remark}

The structure equations for such immersions are
\begin{equation}
 \aligned
   &\partial_j\vec V_i=\varGamma_{ij}^k\vec V_k
    +\varOmega_{ij}^\alpha\vec n_\alpha-Kg_{ij}\vec V, \\
   &\partial_i\vec n_\alpha=-g^{lk}\varOmega_{il}^\alpha\vec V_k,
   \endaligned \label{eq:struct}
\end{equation}
where $\vec V_i=\partial_i\vec V$, $\varGamma_{ij}^k$'s are the
Christofel symbols corresponding to the metric
$(g_{ij})$, $\partial_i=\partial/\partial x_i$. 

From (\ref{eq:ff12}),
\begin{equation}
 g_{ij}=\rho_i^2\delta_{ij}, \quad
   \varOmega_{ij}^\alpha=\rho_i\omega_{i\alpha}\delta_{ij}.
\end{equation}
Then (\ref{eq:struct}) becomes
\begin{equation}
 \aligned
   &\partial_j\vec V_i=\frac{\partial_j\rho_i}{\rho_i}\vec V_i
    +\frac{\partial_i\rho_j}{\rho_j}\vec V_j, \quad (i\ne j), \\
   &\partial_i\vec V_i=
    -\sum_{k\ne i}\frac{\rho_i\partial_k\rho_i}{\rho_k^2}\vec V_k
    +\frac{\partial_i\rho_i}{\rho_i}\vec V_i
    +\rho_i\omega_{i\alpha}\vec n_\alpha
    -K\rho_i^2\vec V, \\
   &\partial_i\vec n_\alpha=-\frac{\omega_{i\alpha}}{\rho_i}\vec V_i. 
   \endaligned \label{eq:structhere}
\end{equation}

The Gauss-Codazzi equations are the integrability conditions for
(\ref{eq:structhere}), which are 
\begin{equation}
 \aligned
   &\partial_i\gamma_{ij}+\partial_j\gamma_{ji}
    =\sum_{l\ne i,\,l\ne j}\gamma_{li}\gamma_{lj},
    \quad (i\ne j), \\
   &\partial_k\gamma_{ij}+\gamma_{ik}\gamma_{kj}=0,\quad 
    (i\ne j,\,i\ne k,\,j\ne k),\\
   &\partial_i\gamma_{ji}+\partial_j\gamma_{ij}
    -\sum_{l\ne i,\,l\ne j}\gamma_{il}\gamma_{jl}-K\rho_i\rho_j=0,
    \quad (i\ne j),\\
   &\partial_j\rho_i+\gamma_{ij}\rho_j=0, \quad(i\ne j), \\
   \endaligned \label{eq:GaussCodazzi}
\end{equation}
\begin{equation}
 \aligned
   &\partial_j\omega_{i\alpha}+\gamma_{ij}\omega_{j\alpha}=0,\quad
    (i\ne j), \\
   &\partial_i\omega_{i\alpha}=\sum_{k\ne i} \gamma_{ki}\omega_{k\alpha},
    \endaligned
   \label{eq:omega}
\end{equation}
where
\begin{equation}
   \gamma_{ij}=\cases -\partial_j\rho_i/\rho_j, &\hbox{if }i\ne j,\\
     0, &\hbox{if }i=j. 
    \endcases
\end{equation}

Let 
\begin{equation}
 \vec p_i=\sum_\alpha\omega_{i\alpha}\vec n_\alpha, \quad
   \vec q_i=\rho_i^{-1}\vec V_i, \quad 
   \vec r=\vec V, \label{eq:defpqr}
\end{equation}
then (\ref{eq:structhere}) becomes
\begin{equation}
 \aligned
   &\partial_i\vec p_i=\sum_{k\ne i}\gamma_{ki}\vec p_k-\vec q_i,\qquad
    \partial_j\vec p_i=-\gamma_{ij}\vec p_j\quad (i\ne j),\\
   &\partial_i\vec q_i=\sum_{k\ne i}\gamma_{ik}\vec q_k+\vec p_i
    -K\rho_i\vec r,\qquad
    \partial_j\vec q_i=-\gamma_{ji}\vec q_j\quad (i\ne j),\\
   &\partial_i\vec r=\rho_i\vec q_i.
   \endaligned \label{eq:pqr}
\end{equation}

For simplicity, we write
\begin{equation}
 \vec p=\pmatrix \vec p_1\\ \vdots\\ \vec p_n \endpmatrix, \quad
   \vec q=\pmatrix \vec q_1\\ \vdots\\ \vec q_n \endpmatrix.
\end{equation}

In \cite{bib:Terng}, a spectral parameter was proposed so that
(\ref{eq:GaussCodazzi}) has a Lax set (generalized Lax pair) 
\begin{equation}
 \aligned
   &\partial_i u_i=\sum_{k\ne i}\gamma_{ki}u_k-\lambda v_i,\qquad
    \partial_j u_i=-\gamma_{ij}u_j\quad (i\ne j), \\
   &\partial_i v_i=\sum_{k\ne i}\gamma_{ik}v_k+\lambda u_i
    -K\rho_i w,\qquad
    \partial_j v_i=-\gamma_{ji}v_j\quad (i\ne j), \\
   &\partial_i w=\rho_i v_i.
   \endaligned \label{eq:lpgeo}
\end{equation}
Note that (\ref{eq:pqr}) is a special case of (\ref{eq:lpgeo}) for
$\lambda=1$. 

Let $\varPhi(\lambda)=(u_1,\cdots,u_n,v_1,\cdots,v_n,w)^T$,
the above system of equations can be written in matrix form as
\begin{equation}
 \partial_i\varPhi=(\lambda J_i+[J_i,P])\varPhi, \label{eq:lpgeomx}
\end{equation}
where
\begin{equation}
 \aligned
   &J_i=\pmatrix 0      &-E_{ii} &0 \\ 
               E_{ii} &0       &0 \\ 
               0      &0       &0 \endpmatrix,\quad
    P=\pmatrix 
     0       &\sum_{i\ne j}\gamma_{ij}E_{ij} &-K\sum_i \rho_ie_i\\
     -\sum_{i\ne j}\gamma_{ji}E_{ij}    &0   &0 \\
     \sum_i \rho_ie_i^T                 &0   &0
    \endpmatrix, \\
   &[J_i,P]=\pmatrix 
    \sum_{j\ne i}\gamma_{ji}(E_{ij}-E_{ji})  &0  &0 \\
    0  &\sum_{j\ne i}\gamma_{ij}(E_{ij}-E_{ji})  &-K\rho_ie_i \\
    0  &\rho_ie_i^T                              &0
   \endpmatrix. 
   \endaligned
\end{equation}
Here $E_{ij}$ is a constant matrix whose $(i,j)$th
entry is one and the rest entries are zero, $e_i$ is a column
matrix whose $i$th entry is one and the rest entries are zero.

If we write a $(2n+1)\times(2n+1)$ matrix $M$ as a block matrix
\vskip-6pt
\begin{equation}
 \matrix 
     \matrix\!\!\phantom{M}\scriptstyle n 
      &\phantom{M_{[2]}}\scriptstyle n 
      &\phantom{M_{[2]}}\scriptstyle 1 \endmatrix 
     &\\
     \left({\matrix M_{[1,1]}&M_{[1,2]}&M_{[1,3]}\\
      M_{[2,1]}&M_{[2,2]}&M_{[2,3]}\\
      M_{[3,1]}&M_{[3,2]}&M_{[3,3]}\endmatrix}\right)
     &\!\!\!\!\!\!\matrix {\scriptstyle n}\phantom M\\ 
     {\scriptstyle n}\phantom M\\ 
     {\scriptstyle 1,}\phantom M\endmatrix
   \endmatrix
\end{equation}
\vskip-6pt
\noindent then
\vskip-6pt
\begin{equation}
 (\rho_i)=P_{[3,1]},\quad
   (\gamma_{ij})=P_{[1,2]},\quad
   (\omega_{ij})=\varPhi_{[1,1]}(0)L_\omega, \quad
   \pmatrix \vec p\\\vec q\\\vec r\endpmatrix=\varPhi(1)\vec L
\end{equation}
\vskip-6pt\noindent 
where $L_\omega$ is an $n\times n$ constant matrix,
$\vec L$ is a $(2n+1)\times 1$ constant matrix.

For the unification of three cases $K=0,\pm1$, we apply
Theorem~\ref{thm2} to (\ref{eq:lpgeomx}). Notice that
(\ref{eq:pqr}) for $(\vec p_i,\vec q_i,\vec r)$ is a special case
of (\ref{eq:lpgeo}) for $(u_i,v_i,w)$ when $\lambda=1$. Moreover,
$\omega_{i\alpha}$'s satisfy the same equations (in
(\ref{eq:omega})) as $u_i$'s do in (\ref{eq:lpgeo}) with 
$\lambda=0$. 

For $K=0,1,-1$, $\lambda J_i+[J_i,P]\in\llll^\sigma(\gggg)$ which
corresponds to
\begin{equation}
 \aligned
   &C=\diag(\underbrace{1,\cdots,1}_n,\underbrace{1,\cdots,1}_n,K), \\
   &\sigma=\diag(\underbrace{1,\cdots,1}_n,\underbrace{-1,\cdots,-1}_n,-1).
   \endaligned
\end{equation}
Hence
\begin{equation}
 C\sigma=\diag(\underbrace{1,\cdots,1}_n,\underbrace{-1,\cdots,-1}_n,-K)
\end{equation}
is always not semidefinite. Therefore, we can use Theorem~\ref{thm2}.

Let $\mu$ be a real number. Let
$H=(\xi_1,\cdots,\xi_n,\I\eta_1,\cdots,
\I\eta_n,\I\zeta)$
be a solution of (\ref{eq:lpgeo}) with $\lambda=\I\mu$ such that
$\sum_i\xi_i^2=\sum_i\eta_i^2+K\zeta^2$. Then $\tau H$ is real and
$H^*C\sigma H=0$. Written explicitly, the
components of $H$ satisfy
\begin{equation}
 \aligned
   &\partial_i \xi_i=\sum_{k\ne i}\gamma_{ki}\xi_k+\mu\eta_i,\qquad
    \partial_j \xi_i=-\gamma_{ij}\xi_j\quad (i\ne j), \\
   &\partial_i \eta_i=\sum_{k\ne i}\gamma_{ik}\eta_k+\mu\xi_i
    -K\rho_i \zeta,\qquad
    \partial_j \eta_i=-\gamma_{ji}\eta_j\quad (i\ne j), \\
   &\partial_i \zeta=\rho_i \eta_i.
   \endaligned \label{eq:lpxi}
\end{equation}
The Darboux matrix in Theorem~\ref{thm2} is
\begin{equation}
 \aligned
   G(\lambda)=&\frac 1{\lambda^2+\mu^2}\left(
    \lambda^2+\mu^2+\frac{2\lambda\mu}{\varDelta}
    \pmatrix 0 &-\xi\eta^T &-K\zeta\xi \\
             \eta\xi^T  &0 &0 \\
             \zeta\xi^T &0 &0 \endpmatrix \right.\\
    &\left.-\frac{2\mu^2}{\varDelta}
    \pmatrix \xi\xi^T &0   &0 \\
             0 &\eta\eta^T  &K\eta\zeta \\
             0 &\zeta\eta^T &K\zeta^2 \endpmatrix\right) 
   \endaligned\label{eq:DM}
\end{equation}
where $\xi=(\xi_1,\cdots,\xi_n)^T$, $\eta=(\eta_1,\cdots,\eta_n)^T$,
$\dsize\varDelta=\sum_{l=1}^n\xi_l^2$. 

The corresponding geometric quantities are given by

\begin{equation}
   \widetilde P
    =P-\frac{2\I\mu}{H^*CH}[\sigma,HH^*]\sigma C,
    \quad \widetilde\omega=G(0)_{[1,1]}\omega,
\end{equation}
\begin{equation}
   (\widetilde{\vec p_1},\cdots,\widetilde{\vec p_n},
   \widetilde{\vec q_1},\cdots,\widetilde{\vec q_n},\widetilde{\vec r})^T
   =G(1)(\vec p_1,\cdots,\vec p_n,\vec q_1,\cdots,\vec q_n,\vec r)^T.
   \label{eq:dtgeo}
\end{equation}

Written explicitly, they are
\begin{equation}
 \aligned
   &\widetilde\rho_i=\rho_i-\frac{2\mu\zeta\xi_i}\varDelta, \quad
    \widetilde\gamma_{ij}=\gamma_{ij}+\frac{2\mu\xi_i\eta_j}\varDelta,
    \quad
    \widetilde\omega_{i\alpha}=\omega_{i\alpha}
    -\sum_j\frac{2\xi_i\xi_j}\varDelta\omega_{j\alpha}, \\
   &\widetilde{\vec p_i}=\vec p_i
    -\sum_j\frac{b\mu\xi_i\xi_j}\varDelta\vec p_j
    -\sum_j\frac{b\xi_i\eta_j}\varDelta\vec q_j
    -\frac{Kb\zeta\xi_i}\varDelta\vec r, \\
   &\widetilde{\vec q_i}=\vec q_i
    +\sum_j\frac{b\eta_i\xi_j}\varDelta\vec p_j
    -\sum_j\frac{b\mu\eta_i\eta_j}\varDelta\vec q_j
    -\frac{Kb\mu\zeta\eta_i}\varDelta\vec r, \\
   &\widetilde{\vec r}=\vec r
    +\sum_j\frac{b\zeta\xi_j}\varDelta\vec p_j
    -\sum_j\frac{b\mu\zeta\eta_j}\varDelta\vec q_j
    -\frac{Kb\mu\zeta^2}\varDelta\vec r,
   \endaligned \label{eq:dtpqr}
\end{equation}
where $\dsize b=\frac{2\mu}{1+\mu^2}$.

Using (\ref{eq:defpqr}) and (\ref{eq:structhere}), we have
\begin{equation}
 \aligned
   \widetilde{\vec V}=&\sum_j\frac{b\zeta\xi_j}{\varDelta\rho_j}
    \frac{\partial^2\vec V}{\partial x_j^2} 
    +\sum_j\frac{b\zeta}{\varDelta\rho_j^2}\big(
    -\xi_j\partial_j\rho_j+\sum_{k\ne j}\xi_k\partial_j\rho_k
    -\mu\rho_j\eta_j\big)\frac{\partial\vec V}{\partial x_j} \\
   &+\left(1-\frac{Kb\mu\zeta^2}\varDelta
    +\sum_j\frac{Kb\zeta\rho_j\xi_j}\varDelta\right)\vec V.
   \endaligned \label{eq:dtV}
\end{equation}

Therefore, we have the following general procedure to get the
local isometric immersion $M_n(K)\to M_{2n}(K)$ with flat normal
bundle and linearly independent curvature normals. Since we
consider the linear system (\ref{eq:lpgeo}), the condition
$\rho_i\ne 0$ can be removed temporarily, provided that 
$\widetilde\rho_i\ne 0$ for derived submanifold.
\item{1)} Suppose we know a solution of (\ref{eq:GaussCodazzi})
and can solve the linear system (\ref{eq:lpgeo}) to get the
fundamental solution $\varPhi(x_1,\cdots,x_n,\lambda)$. 
\item{2)} Let $\lambda=\I\mu$ and get a solution
$(\xi,\eta,\zeta)$ of (\ref{eq:lpxi}) such that
\begin{equation}
 (\xi,\I\eta,\I\zeta)^T
   =\varPhi(x_1,\cdots,x_n,\I\mu)C
\end{equation}
where $C$ is a constant matrix.
\item{3)} Using (\ref{eq:dtpqr}), one gets the position vector
$\widetilde{\vec r}$ and the corresponding quantities. When
$\rho_i\ne 0$, (\ref{eq:dtV}) gives a more direct answer.
Moreover, the corresponding solution of (\ref{eq:lpgeo}) is
$\widetilde\varPhi(x_1,\cdots,x_n,\lambda)
=G(x_1,\cdots,x_n,\lambda)\varPhi(x_1,\cdots,x_n,\lambda)$.

These three steps give the new solution of
(\ref{eq:GaussCodazzi}), (\ref{eq:pqr}). When 
$\widetilde\rho_i\ne 0$, the solution represents a real geometric
immersion. 

For $\widetilde{\vec r}$, the corresponding solution of
(\ref{eq:lpgeo}) is known. Hence, we can repeat step~2) and~3) to
get another solution $\widetilde{\widetilde{\vec r}}$. Continuing
this process, a series of immersions are obtained by an algebraic
algorithm and one should solve a system of linear ODEs only once
in step~1). 

Now we give some examples. The simplest examples are the totally
geodesic submanifolds. However, since their second fundamental
forms are zero, the  Darboux transformation here is not
applicable (also, their curvature normals are linearly
dependent). Therefore, we seek other solutions as seed
solutions of Darboux transformation. 
\medskip
\noindent{\bf (1) Solutions derived from the trivial solutions}
\smallskip
We take the seed solution of (\ref{eq:GaussCodazzi}) as $\rho_i=0$,
$\gamma_{ij}=0$, $\omega_{i\alpha}=\delta_{i\alpha}$. Although
this does not correspond to a real geometric object, and
(\ref{eq:dtV}) is not valid, we can still get 
$(\widetilde{\vec p},\widetilde{\vec q},\widetilde{\vec r})$ 
from (\ref{eq:dtpqr}) and then get the nontrivial local immersion
$\widetilde{\vec r}$. 

To get a solution, we first solve (\ref{eq:pqr}) to get 
$(\vec p,\vec q,\vec r)$ and solve (\ref{eq:lpxi}) to get
$(\xi_i,\eta_i,\zeta)$. Then normalize the first fundamental form
by changing coordinates $\{x_i\}$. After that, the last equation
of (\ref{eq:dtpqr}) gives the position vector of the immersion.

First, we solve (\ref{eq:pqr}) to get
\begin{equation}
 \aligned
   &\vec p_i=\vec E_i\cos x_i+\vec F_i\sin x_i, \\
   &\vec q_i=\vec E_i\sin x_i-\vec F_i\cos x_i, \\
   &\vec r=\vec R,
   \endaligned
\end{equation}
where $\vec E_i$, $\vec F_i$, $\vec R$ $(i=1,\cdots,n)$ are
constant vectors in $\hr_{2n+1}(K)$ such that
\begin{equation}
 \aligned
   &\vec E_i\cdot\vec E_j=\delta_{ij},\quad
   \vec E_i\cdot\vec F_j=0,\quad
   \vec F_i\cdot\vec F_j=\delta_{ij},\quad
   \vec E_i\cdot\vec R=\vec F_i\cdot\vec R=0, \\
   &\vec R\cdot\vec R=K \; (\hbox{if }K=\pm 1).
   \endaligned
\end{equation}

A) $K=0$. \par
The solution of (\ref{eq:lpxi}) is
\begin{equation}
 \xi_i=A_ie^{\mu x_i}, \quad 
   \eta_i=A_ie^{\mu x_i}, \quad \zeta=C
\end{equation}
where $A_i$, $C$ are nonzero real constants.

From (\ref{eq:dtpqr}),
\begin{equation}
 \widetilde\rho_i
   =-\frac{2\mu CA_ie^{\mu x_i}}{\sum_k A_k^2e^{2\mu x_k}}.
\end{equation}

Let 
\begin{equation}
   z_i=\frac{2CA_ie^{\mu x_i}}{\sum_k A_k^2e^{2\mu x_k}},
   \label{eq:eg0trans}
\end{equation}
then
\begin{equation}
 \widetilde I=\sum_i\widetilde\rho_i^2\,dx_i^2=\sum_i dz_i^2.
\end{equation}
Choose $\vec R=0$, then
\begin{equation}
 \aligned
   \widetilde{\vec r}&=\sum_i \frac{\mu}{1+\mu^2}
    z_i\left(\vec E_i\cos x_i+\vec F_i\sin x_i\right)
    -\sum_i \frac{\mu^2}{1+\mu^2}
    z_i\left(\vec E_i\sin x_i-\vec F_i\cos x_i\right) \\
   &=\sum_i\frac{\mu z_i}{\sqrt{1+\mu^2}}
    \left(\vec E_i\cos\widetilde x_i+\vec F_i\sin\widetilde x_i\right) 
   \endaligned
\end{equation}
where
\begin{equation}
 \widetilde x_i=x_i+\tg^{-1}\mu=\frac 1\mu\ln\left(\frac {2C}{A_i}
    \frac{z_i}{\sum_k z_k^2}\right)+\tg^{-1}\mu.
\end{equation}
This local immersion is only defined for $(z_1,\cdots,z_n)$ with
$A_iCz_i>0$. 

For $n=2$ and $A_i=1$, $C>0$, let $z_1=r\cos\phi$, $z_2=r\sin\phi$,
$$
   \aligned
   \vec k_1=\frac{\vec E_1+\mu\vec F_1}{\sqrt{1+\mu^2}},\quad
   \vec k_2=\frac{\vec F_1-\mu\vec E_1}{\sqrt{1+\mu^2}},\\
   \vec k_3=\frac{\vec E_2+\mu\vec F_2}{\sqrt{1+\mu^2}},\quad
   \vec k_4=\frac{\vec F_2-\mu\vec E_2}{\sqrt{1+\mu^2}},
   \endaligned
$$
then $\{\vec k_1,\vec k_2,\vec k_3,\vec k_4\}$ is also an 
orthonormal frame of $\hr^4$, and
\begin{equation}
 \aligned
   \widetilde{\vec r}&=\frac{\mu}{\sqrt{1+\mu^2}}\left(
    r\cos\phi\cos\bigg(\frac1\mu\ln\big(\frac{2C}{r}
     \cos\phi\big)\bigg)\vec k_1 
     +r\cos\phi\sin\bigg(\frac1\mu\ln\big(\frac{2C}{r}
     \cos\phi\big)\bigg)\vec k_2 \right. \\
    &\left.+r\sin\phi\cos\bigg(\frac1\mu\ln\big(\frac{2C}{r}
     \sin\phi\big)\bigg)\vec k_3 
     +r\sin\phi\sin\bigg(\frac1\mu\ln\big(\frac{2C}{r}
     \sin\phi\big)\bigg)\vec k_4 \right)
     \label{eq:eg02-1} \\
    &(r>0, 0<\phi<\frac{\pi}2)
   \endaligned
\end{equation}
Written in terms of the coordinates $(x_1,x_2)$,
\begin{equation}
 \aligned
   \widetilde{\vec r}&=\frac{2\mu C}{\sqrt{1+\mu^2}}\frac 1{e^{2\mu x_1}
    +e^{2\mu x_2}}
    \big(e^{\mu x_1}\cos x_1\;\vec k_1+e^{\mu x_1}\sin x_1\;\vec k_2\\
   &+e^{\mu x_2}\cos x_2\;\vec k_3+e^{\mu x_2}\sin x_2\;\vec k_4\big).
    \label{eq:eg02-2}
   \endaligned
\end{equation}

\begin{remark}
When $\mu C$ is finite and $\mu\to 0$, except for $(0,0,0,0)$,
each point on the submanifold (\ref{eq:eg02-2}) tends to the
standard torus in $\hr^4${\rm :}
\begin{equation}
 \vec r=\mu C(\cos x_1\;\vec k_1+\sin x_1\;\vec k_2
   +\cos x_2\;\vec k_3+\sin x_2\;\vec k_4).
\end{equation} 
The coefficient is $\mu C$ because the metric on the surface is 
$$ I=\mu^2 C^2\sum_{i=1}^2 dx_i^2. 
$$
When $C$ is finite and $\mu\to \infty$, the pointwise limit of
the submanifold given by (\ref{eq:eg02-1}) is 
\begin{equation}
 \vec r=r\cos\phi\;\vec k_1+r\sin\phi\;\vec k_3,
\end{equation}
which is a plain in $\hr^4$.
\end{remark}

\begin{remark}
In the coordinate $\{z_i\}$, (\ref{eq:pqr}) or
(\ref{eq:structhere}) is no longer satisfied, because those
equations depend on the special coordinate $\{x_i\}$.
\end{remark}

The following figures show the manifold with $\mu=-0.2$, $C=1$. 
In Fig.~1, the three axes are $(r_1,r_2,r_3)$
if $\vec r$ is written as $\vec r=\sum_{j=1}^4 r_j\;\vec k_j$. 
Fig.~2 is the corresponding contour plot of Fig.~1 on $(r_1,r_2)$
plane. In Fig.~3, the axes are $(r_1,r_2,\pm\sqrt{r_3^2+r_4^2})$.
The dark spiral in Fig.~2 represents the boundary of the manifold.

\medskip
\unitlength=1mm
\ifx\figuretype\BMPtype
{
\hbox{\vbox{\divide\hsize by 2\advance\hsize by -0.6cm
\begin{picture}(38,27)(-10,-27)
\put(0,0){\special{em:graph FIG1.BMP}}
\end{picture}
\smallskip
\null\hskip3cm{\small Fig.~1}
}
\vbox{\divide\hsize by 2\advance\hsize by -0.6cm
\begin{picture}(38,33)(-10,-33)
\put(0,0){\special{em:graph FIG2.BMP}}
\end{picture}
\smallskip
\null\hskip3.5cm{\small Fig.~2}
}\hfill}
\vskip-6pt
\medskip\vskip0pt plus 1fill
}
\else
{
\vskip-2truecm
\epsffile[-60 0 402 261]{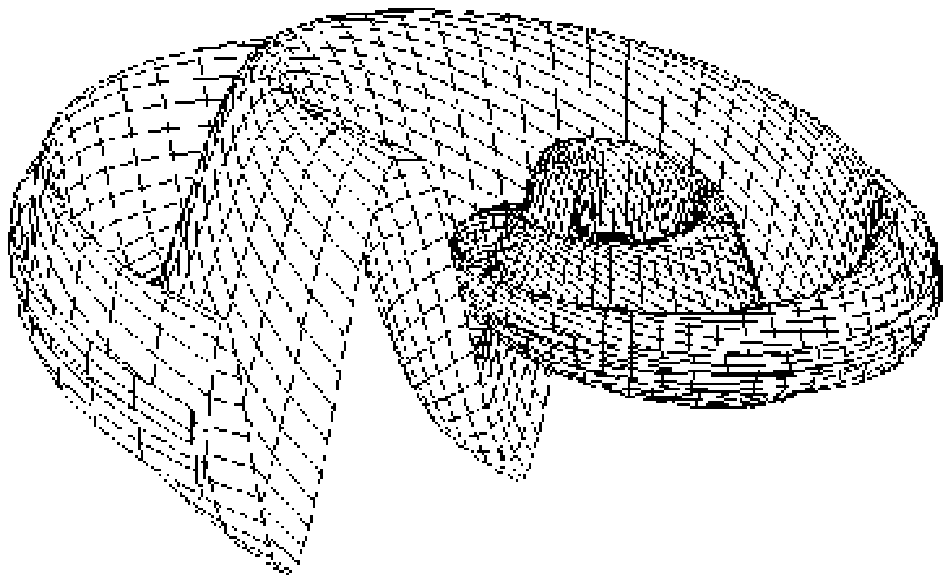}
\hbox to \hsize{\hfill Fig.~1 \hfill}
\vskip-2truecm
\epsffile[-60 0 402 286]{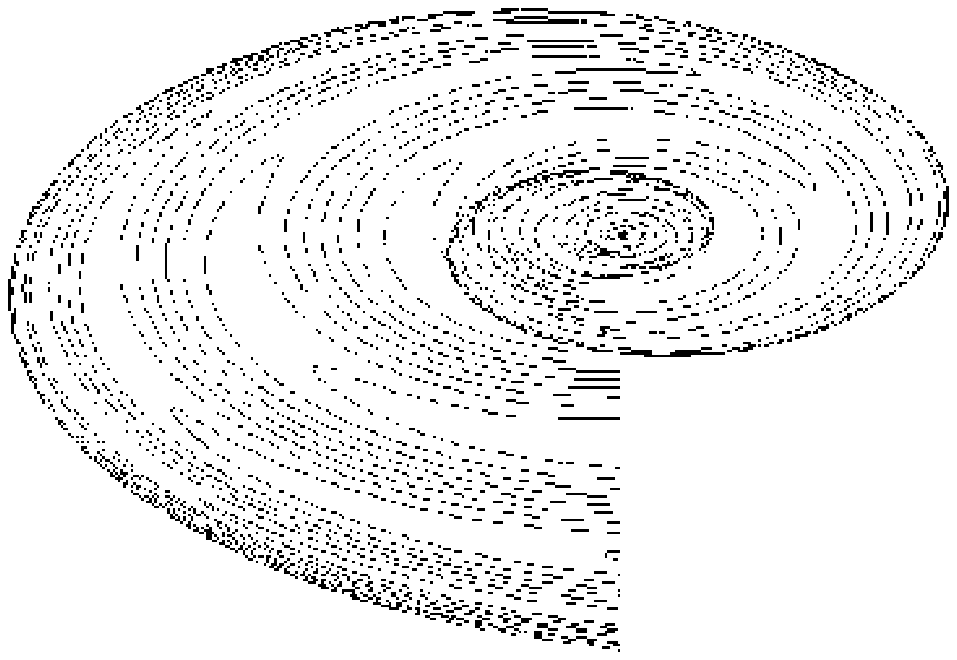}
\hbox to \hsize{\hfill Fig.~2 \hfill}
}
\fi

\if\figuretype\BMPtype
{
\vbox{\divide\hsize by 2\advance\hsize by -0.6cm
\begin{picture}(38,25)(-10,-25)
\put(0,0){\special{em:graph FIG3.BMP}}
\end{picture}
\smallskip
\null\hskip3cm{\small Fig.~3}
}
\medskip\vskip0pt plus 1fill
}
\else
{
\vskip-2truecm
\epsffile[-60 0 402 259]{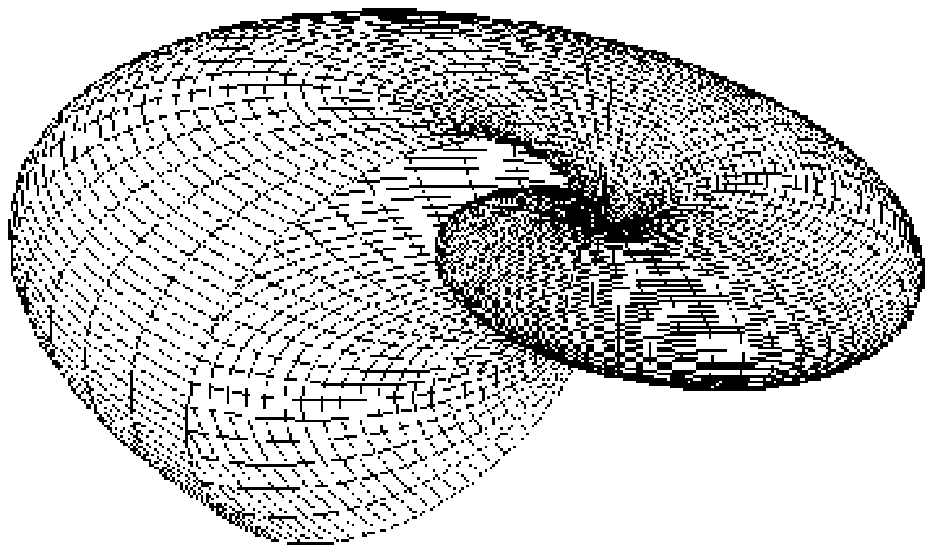}
\hbox to \hsize{\hfill Fig.~3 \hfill}
}
\fi

B) $K=1$.\par
The solution of (\ref{eq:lpxi}) is
\begin{equation}
 \xi_i=A_ie^{\mu x_i}+B_ie^{-\mu x_i}, \quad 
   \eta_i=A_ie^{\mu x_i}-B_ie^{-\mu x_i}, \quad \zeta=C
\end{equation}
where $A_i$, $B_i$, $C$ are nonzero real constants with
$C^2=4\sum_i A_iB_i$. 

The corresponding
\begin{equation}
 \widetilde\rho_i=-\frac
    {2\mu C(A_ie^{\mu x_i}+B_ie^{-\mu x_i})}
    {\sum_k(A_ke^{\mu x_k}+B_ke^{-\mu x_k})^2}.
\end{equation}
Let
\begin{equation}
 z_i=\frac{A_ie^{\mu x_i}-B_ie^{-\mu x_i}}{C},
\end{equation}
then
\begin{equation}
 \widetilde I=\sum_i\widetilde\rho_i^2\,dx_i^2
   =\frac{4\sum_idz_i^2}{(\sum_k z_k^2+1)^2},
\end{equation}
\begin{equation}
 \aligned
   \widetilde{\vec r}&=\left(1-\frac{b\mu}{\sum_k z_k^2+1}\right)\vec R
    +\frac{b\sqrt{z_i^2+C_i}}{\sum_k z_k^2+1}
    (\vec E_i\cos x_i+\vec F_i\sin x_i) \\
   &\quad-\frac{b\mu z_i}{\sum_k z_k^2+1}
    (\vec E_i\sin x_i-\vec F_i\cos x_i) \\
   &=\left(1-\frac{2\mu^2}{1+\mu^2}\frac1{\sum_k z_k^2+1}\right)\vec R
    +\frac{2\mu\sqrt{(1+\mu^2)z_i^2+C_i}}{(1+\mu^2)(\sum_k z_k^2+1)}
    (\vec E_i\cos\widetilde x_i+\vec F_i\sin\widetilde x_i)
   \endaligned
\end{equation}
with $C_i=4A_iB_i/C^2$ (therefore $\sum_i C_i=1$),
\begin{equation}
 \widetilde x_i=\frac 1\mu\ln
    \left(\frac{C}{2A_i}(z_i+\sqrt{z_i^2+C_i})\right)
    +\tg^{-1}\frac{\mu z_i}{\sqrt{z_i^2+C_i}}.
\end{equation}

It is clear that this immersion is smooth for $|z|<\infty$. It
can be verified that at $|z|=\infty$ (by changing coordinates) it
is not smooth. 

C) $K=-1$. \par
The solution of (\ref{eq:lpxi}) is
\begin{equation}
 \xi_i=A_ie^{\mu x_i}-B_ie^{-\mu x_i}, \quad 
   \eta_i=A_ie^{\mu x_i}+B_ie^{-\mu x_i}, \quad \zeta=C
\end{equation}
where $A_i$, $B_i$, $C$ are nonzero real constants with
$C^2=4\sum_i A_iB_i$. 
Then
\begin{equation}
 \widetilde\rho_i=-\frac
    {2\mu C(A_ie^{\mu x_i}-B_ie^{-\mu x_i})}
    {\sum_k(A_ke^{\mu x_k}-B_ke^{-\mu x_k})^2}.
\end{equation}

Let
\begin{equation}
 z_i=\frac{C(A_ie^{\mu x_i}+B_ie^{-\mu x_i})}
   {\sum_k(A_ke^{\mu x_k}+B_ke^{-\mu x_k})^2},
\end{equation}
then
\begin{equation}
 \widetilde I=\sum_i\widetilde\rho_i^2\,dx_i^2
   =\frac{4\sum_idz_i^2}{(1-\sum_k z_k^2)^2},
\end{equation}
\begin{equation}
 \aligned
   \widetilde{\vec r}&=\left(1+\frac{2\mu^2}{1+\mu^2}
    \frac{\sum_k z_k^2}{1-\sum_k z_k^2}\right)\vec R
    +\frac{2\mu\sqrt{(1+\mu^2)z_i^2-C_i(\sum_k z_k^2)^2}}
     {(1+\mu^2)(1-\sum_k z_k^2)}
    (\vec E_i\cos\widetilde x_i+\vec F_i\sin\widetilde x_i)
   \endaligned
   \label{eq:vec-1}
\end{equation}
with $C_i=4A_iB_i/C^2$ (also $\sum_i C_i=1$),
\begin{equation}
 \widetilde x_i=\frac 1\mu\ln
    \left(\frac C{2A_i}
    \frac{z_i+\sqrt{z_i^2-C_i(\sum_k z_k^2)^2}}{\sum_k z_k^2}\right)
    +\tg^{-1}\frac{\mu z_i}{\sqrt{z_i^2-C_i(\sum_k z_k^2)^2}}.
\end{equation}

Now $M_n(-1)$ is the disk $\sum_k z_k^2<1$, but the immersion
(\ref{eq:vec-1}) can only be defined in a region of $M_n(-1)$.
For example, when all $A_i>0$, $B_i>0$, it is defined in
$C_i(\sum_k z_k^2)^2<z_i^2$, which contains the center
$(0,0,\cdots,0)$ of the disk.

\medskip
\noindent{\bf (2) Solutions derived from the torus $T^n\to\hr^{2n}$}
\smallskip
Let
\begin{equation}
 \vec r=\sum_{i=1}^n\left(
    \cos x_i\vec E_i+\sin x_i\vec F_i\right),
\end{equation}
where $\vec E_i$, $\vec F_i$ are constant vectors satisfying
\begin{equation}
 \vec E_i\cdot\vec E_j=\delta_{ij},\quad
   \vec E_i\cdot\vec F_j=0,\quad
   \vec F_i\cdot\vec F_j=\delta_{ij}.
\end{equation}
This is the standard torus $T^n$ in $\hr^{2n}$, whose
$g_{ij}=\delta_{ij}$, $\omega_{i\alpha}=\delta_{i\alpha}$, and
the corresponding 
\begin{equation}
 \aligned
   &\vec p_i=-\cos x_i\vec E_i-\sin x_i\vec F_i, \\
   &\vec q_i=-\sin x_i\vec E_i+\cos x_i\vec F_i.
   \endaligned
\end{equation}
Solving (\ref{eq:lpxi}) with $\sum_i\xi_i^2=\sum_i\eta_i^2$, we have
\begin{equation}
 \xi_i=A_ie^{\mu x_i}, \quad 
   \eta_i=A_ie^{\mu x_i}, \quad 
   \zeta=\sum_{k=1}^n\frac{A_k}{\mu}e^{\mu x_k}+C,
\end{equation}
where $A_i$, $C$ are real constants. Then (\ref{eq:dtpqr}) gives
\begin{equation}
   \aligned
   \widetilde{\vec r}=&\sum_{i=1}^n(\cos x_i\vec E_i+\sin x_i\vec F_i)\\
   &-\frac b\varDelta\big(\sum_{k=1}^n\frac{A_k}\mu e^{\mu x_k}+C\big)
    \sum_{i=1}^nA_ie^{\mu x_i}\big((\cos x_i-\mu\sin x_i)\vec E_i
    +(\sin x_i+\mu\cos x_i)\vec F_i\big),
   \endaligned
\end{equation}
\begin{equation}
   \widetilde\rho_i=1-\frac{2A_ie^{\mu x_i}}{\varDelta}
   \Big(\sum_{k=1}^nA_ke^{\mu x_k}+\mu C\Big)
\end{equation}
where
\begin{equation}
 \varDelta=\sum_kA_k^2e^{2\mu x_k},\quad b=\frac{2\mu}{1+\mu^2}.
\end{equation}
This submanifold is defined in the region where
$\widetilde\rho_i\ne 0$ for all $1\le i\le n$.

By the local change of coordinates
\begin{equation}
   z_i=x_i-\frac{2A_ie^{\mu x_i}}{\mu\varDelta}
   \Big(\sum_{k=1}^nA_ke^{\mu x_k}+\mu C\Big),
\end{equation}
the metric is changed to 
$$ \widetilde I=\sum_{i=1}^n dz_i^2. $$

Fig.~4 shows this submanifold (surface) for $n=2$, $A_i=1$, $C=1$,
$\mu=-0.2$. Write $\vec r=r_1\vec E_1+r_2\vec F_1+r_3\vec
E_2+r_4\vec F_2$, then the coordinates are taken as
$\D((2+r_1)r_3,(2+r_1)r_4,r_2)$. Fig.~5 is the corresponding
figure for the standard torus.

\medskip
\unitlength=1mm
\ifx\figuretype\BMPtype
{
\hbox{\vbox{\divide\hsize by 2\advance\hsize by -0.6cm
\begin{picture}(38,33)(-10,-33)
\put(0,0){\special{em:graph FIG4.BMP}}
\end{picture}
\smallskip
\null\hskip3cm{\small Fig.~4}
}
\vbox{\divide\hsize by 2\advance\hsize by -0.6cm
\begin{picture}(38,33)(-10,-33)
\put(0,0){\special{em:graph FIG5.BMP}}
\end{picture}
\smallskip
\null\hskip3.5cm{\small Fig.~5}
}\hfill}
\vskip1cm
}
\else
{
\vskip-2truecm
\epsffile[-60 0 402 282]{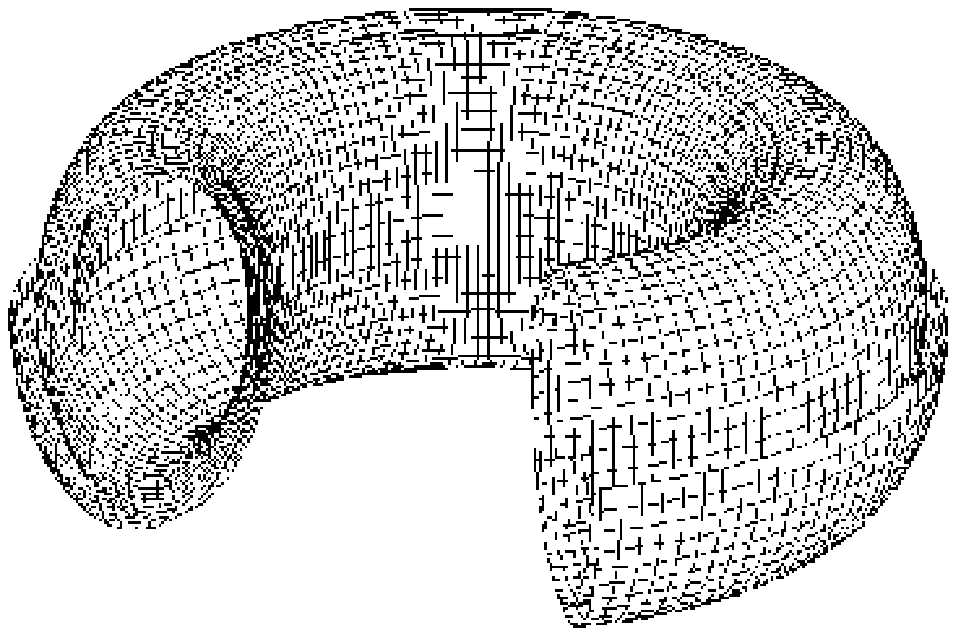}
\hbox to \hsize{\hfill Fig.~4 \hfill}
\vskip-2truecm
\epsffile[-60 0 402 271]{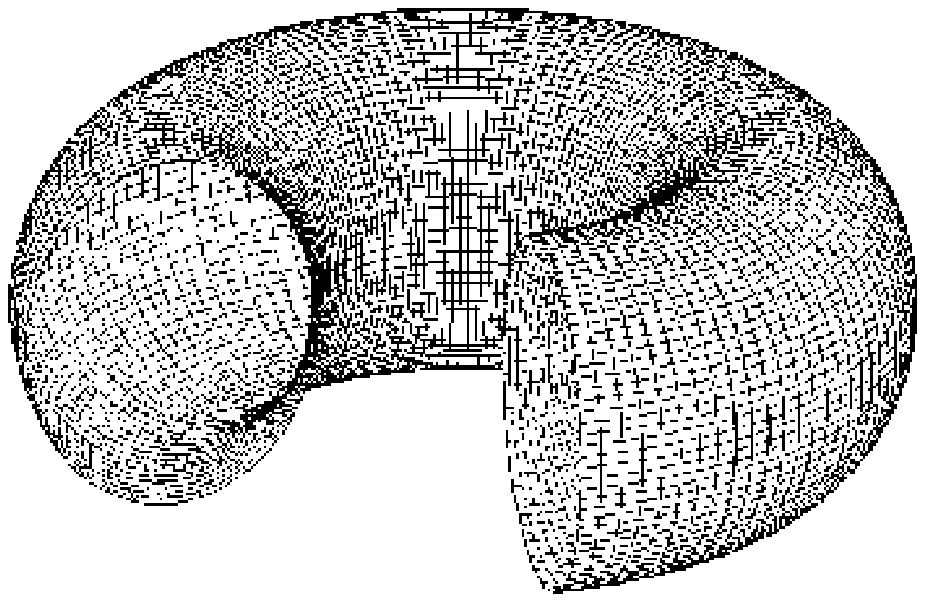}
\hbox to \hsize{\hfill Fig.~5 \hfill}
}
\fi
\noindent{\bf Acknowledgements}\par
The paper was supported by Chinese National Research Project
``Nonlinear Science'', Chinese National Science Foundation for
Youth, and Fok~Ying-Tung Education Foundation of China.
The author would like to thank Prof.~C.~H.~Gu, Prof.~H.~S.~Hu, 
Prof.~Y.~L.~Xin, Prof.~J.~X.~Hong, Prof.~X.~G.~Huang,
Dr.~J.~M.~Lin and Dr.~Q.~Ding for encouragements and helpful
discussions.

\end{document}